      [1999/12/01 v1.4c Il Nuovo Cimento]
\documentclass{cimento}

\def\dd{\displaystyle}

\def\bea{\begin{eqnarray}}
\def\eea{\end{eqnarray}}
\def\beq{\begin{equation}}
\def\eeq{\end{equation}}
\def\bq{\begin{quote}}
\def\eq{\end{quote}}
\def\be{\begin{equation}}
\def\ee{\end{equation}}
\def\bc{\begin{center}}
\def\ec{\end{center}}
\def\bea{\begin{eqnarray}}
\def\eea{\end{eqnarray}}
\def\dd{\displaystyle}

\def\GeV{{\rm GeV}}
\def\gappeq{\mathrel{\rlap {\raise.5ex\hbox{$>$}} {\lower.5ex\hbox{$\sim$}}}}
\def\lappeq{\mathrel{\rlap{\raise.5ex\hbox{$<$}} {\lower.5ex\hbox{$\sim$}}}}

\usepackage{graphicx}  

\title{Status of Neutrino Masses and Mixing in 2009}
\author{Guido~Altarelli\from{ins:x}}
\instlist{\inst{ins:x} 
Dipartimento di Fisica `E.~Amaldi', Universit\`a di Roma Tre
and INFN, Sezione di Roma Tre, I-00146 Rome, Italy and \\ CERN, Department of Physics, Theory Unit, 
 CH--1211 Geneva 23, Switzerland}

\PACSes{
\PACSit{11.30,12.10,12.60,13.15}{\ldots}}
\begin{document}

\maketitle

\begin{abstract}
We present a very concise summary of the status of our knowledge and understanding of neutrino masses and mixing.
\end{abstract}

\begin{flushright}
{RM3-TH/09-10}~~~~~~
{CERN-PH-TH/2009-068}\\
\end{flushright}

\section{Experimental data}

That neutrinos have a mass has been established by experiments on neutrino oscillations that measure differences of squared masses and mixing angles \cite{review}. Two distinct oscillation frequencies have been at first measured in solar and atmospheric neutrino oscillations and later confirmed by experiments on earth, like KamLAND and K2K. A signal corresponding to a third mass difference was claimed by the LSND experiment but not confirmed by KARMEN and recently by MiniBooNE. Two well separated differences need at least three neutrino mass eigenstates involved in oscillations. Conversely the three known neutrino species can be sufficient. At least two $\nu$'s must be massive while, in principle, the third one could still be massless. In the following we will assume the simplest picture with three active neutrinos (CPT invariance and no sterile neutrinos). The mass eigenstates involved in solar oscillations are $m_1$ and $m_2$ and, by definition, $|m_2|> |m_1|$, so that $\Delta m^2_{sun}=|m_2|^2-|m_1|^2>0$. The atmospheric neutrino oscillations involve $m_3$:  $\Delta m^2_{atm}=|\Delta m^2_{31}|$ with $\Delta m^2_{31}=|m_3|^2-|m_1|^2$ either positive (normal hierarchy) or negative (inverse hierarchy). The present data are compatible with both cases. The degenerate spectrum occurs when the average absolute value of the masses is much larger than all mass squared differences: $|m_i|^2 >> \Delta m^2_{hk}$. With the standard set of notations and definitions \cite{review} the present data are summarised in Table 1.

\begin{table}
  \caption{Fit to neutrino oscillation data}
  \label{tab:data}
  \begin{center}
\begin{narrowtabular}{2cm}{rcl}
\hline
&&\\[-4mm]
  & ref. \cite{FogliIndication} & ref. \cite{MaltoniIndication}   \\[2mm]
\hline
& & \\[-4mm]
$(\Delta m^2_{sun})~(10^{-5}~{\rm eV}^2)$ &$7.67^{+0.16}_{-0.19}$ & $7.65^{+0.023}_{-0.020}$\\[2mm]
\hline
& & \\[-4mm]
$\Delta m^2_{atm}~(10^{-3}~{\rm eV}^2)$ &$2.39^{+0.11}_{-0.08}$ & $2.40^{+0.012}_{-0.011}$\\[2mm]
\hline
&&\\[-4mm]
$\sin^2\theta_{12}$ &$0.312^{+0.019}_{-0.018}$ & $0.304^{+0.022}_{-0.016}$\\[2mm]
\hline
&&\\[-4mm]
$\sin^2\theta_{23}$ &$0.466^{+0.073}_{-0.058}$ &  $0.50^{+0.07}_{-0.06}$\\[2mm]
\hline
&&\\[-4mm]
$\sin^2\theta_{13}$ &$0.016\pm0.010$ &$0.010^{+0.016}_{-0.011}$ \\[2mm]
\hline
  \end{narrowtabular}
\end{center}
\end{table}

Oscillation experiments only measure differences of  squared masses  and do not provide information about the absolute neutrino mass scale. Limits on that are obtained \cite{review} from the endpoint of the tritium beta decay spectrum, from cosmology and from neutrinoless double beta decay ($0\nu \beta \beta$). From tritium we have an absolute upper limit of
2.2 eV (at 95\% C.L.) on the mass of electron  antineutrino, which, combined with the observed oscillation
frequencies under the assumption of three CPT-invariant light neutrinos, represents also an upper bound on the masses of
the other active neutrinos. Complementary information on the sum of neutrino masses is also provided by the galaxy power
spectrum combined with measurements of the cosmic  microwave background anisotropies. According to recent analyses of the most reliable data \cite{fo}
$\sum_i \vert m_i\vert < 0.60\div 0.75$ eV (at 95\% C.L.) depending on the retained data (the numbers for the sum have to be divided by 3 in order to obtain a limit on the mass of each neutrino).
The discovery of $0\nu \beta \beta$ decay would be very important because it would establish lepton number violation and
the Majorana nature of $\nu$'s, and provide direct information on the absolute
scale of neutrino masses.
As already mentioned the present limit from $0\nu \beta \beta$  (with large ambiguities from nuclear matrix elements) is about $\vert m_{ee}\vert < (0.3\div 0.8)$ eV \cite{fo} (see eq. (\ref{3nu1gen}). 

\section{Majorana Neutrinos and the See-Saw Mechanism}

Given that neutrino masses are certainly extremely
small, it is really difficult from the theory point of view to avoid the conclusion that the lepton number L conservation is probably violated and that $\nu$'s are Majorana fermions.
In this case the smallness of neutrino masses can be naturally explained as inversely proportional
to the very large scale where L is violated, of order the grand unification scale $M_{GUT}$ or maybe, for the lightest among them, the Planck scale $M_{Pl}$. 
If neutrinos are Majorana particles, their masses arise  from the generic dimension-five non renormalizable operator of the form: 
\be
O_5=\frac{(H l)^T_i \lambda_{ij} (H l)_j}{M}+~h.c.~~~,
\label{O5}
\ee  
with $H$ being the ordinary Higgs doublet, $l_i$ the SU(2) lepton doublets, $\lambda$ a matrix in  flavour space,
$M$ a large scale of mass and a charge conjugation matrix $C$
between the lepton fields is understood. 

Neutrino masses generated by $O_5$ are of the order
$m_{\nu}\approx v^2/M$ for $\lambda_{ij}\approx {\rm O}(1)$, where $v\sim {\rm O}(100~\GeV)$ is the vacuum
expectation value of the ordinary Higgs. A particular realization leading to comparable masses is the see-saw mechanism \cite{seesaw}, where $M$ derives from the exchange of heavy neutral objects of weak isospin 0 or 1. In the simplest case the exchanged particle is the $\nu_R$ and the resulting neutrino mass matrix reads (1st type see-saw ):
\be  
m_{\nu}=m_D^T M^{-1}m_D~~~.
\ee 
As one sees,  the light neutrino masses are quadratic in the Dirac
masses and inversely proportional to the large Majorana mass.  For
$m_{\nu}\approx \sqrt{\Delta m^2_{atm}}\approx 0.05$ eV and 
$m_{\nu}\approx m_D^2/M$ with $m_D\approx v
\approx 200$~GeV we find $M\approx 10^{15}$~GeV which indeed is an impressive indication that the scale for lepton number violation is close to
$M_{GUT}$. Thus probably neutrino masses are a probe into the physics near $M_{GUT}$. This argument, in my opinion, strongly discourages models where neutrino masses are generated near the weak scale and are suppressed by some special mechanism.

\section{Importance of Neutrinoless Double Beta Decay}

Oscillation experiments cannot distinguish between
Dirac and Majorana neutrinos.
The detection of neutrino-less double beta decay would provide direct evidence of $L$ non conservation, and the Majorana nature of neutrinos. It would also offer a way to possibly disentangle the 3 cases of degenerate, normal or inverse hierachy neutrino spectrum.  The quantity which is bound by experiments on $0\nu \beta \beta$
is the 11 entry of the
$\nu$ mass matrix, which in general, from $m_{\nu}=U^* m_{diag} U^\dagger$, is given by :
\bea 
\vert m_{ee}\vert~=\vert(1-s^2_{13})~(m_1 c^2_{12}~+~m_2 s^2_{12})+m_3 e^{2 i\phi} s^2_{13}\vert
\label{3nu1gen}
\eea
where $m_{1,2}$ are complex masses (including Majorana phases) while $m_3$ can be taken as real and positive and $\phi$ is the $U_{PMNS}$ phase measurable from CP violation in oscillation experiments. Starting from this general formula it is simple to
derive the bounds for degenerate, inverse hierarchy or normal hierarchy mass patterns shown in Fig.1.

\begin{figure}
\centering
\includegraphics [width=10.0 cm]{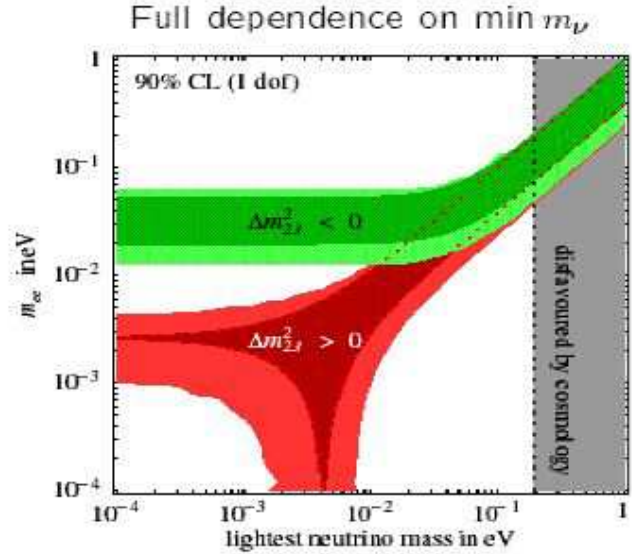}    
\caption{A plot \cite{fsv} of $m_{ee}$ in eV, the quantity measured in neutrino-less double beta decay, given in eq.(\ref{3nu1gen}), versus the lightest neutrino mass $m_1$, also in eV. The upper (lower) band is for inverse (normal) hierarchy.}
\end{figure}

In the next few years a new generation of experiments will reach a larger sensitivity on $0\nu \beta \beta$ by about an order of magnitude. If these experiments will observe a signal this would indicate that the inverse hierarchy is realized, if not, then the normal hierarchy case remains a possibility. 

\section{Baryogenesis via Leptogenesis from Heavy $\nu^c$ Decay}

In the Universe we observe an apparent excess of baryons over antibaryons. It is appealing that one can explain the
observed baryon asymmetry by dynamical evolution (baryogenesis) starting from an initial state of the Universe with zero
baryon number.  For baryogenesis one needs the three famous Sakharov conditions: B violation, CP violation and no thermal
equilibrium. In the history of the Universe these necessary requirements have possibly occurred at different epochs. Note
however that the asymmetry generated by one epoch could be erased at following epochs if not protected by some dynamical
reason. In principle these conditions could be verified in the SM at the electroweak phase transition. B is violated by
instantons when kT is of the order of the weak scale (but B-L is conserved), CP is violated by the CKM phase and
sufficiently marked out-of- equilibrium conditions could be realized during the electroweak phase transition. So the
conditions for baryogenesis  at the weak scale in the SM superficially appear to be present. However, a more quantitative
analysis
\cite{tro} shows that baryogenesis is not possible in the SM because there is not enough CP violation and the phase
transition is not sufficiently strong first order, unless the Higgs mass is below a bound which by now is excluded by LEP. In SUSY extensions of the SM, in particular in the MSSM,
there are additional sources of CP violation and the bound on $m_H$ is modified but also this possibility has by now become at best marginal after the results from LEP2.

If baryogenesis at the weak scale is excluded by the data it can occur at or just below the GUT scale, after inflation.
But only that part with
$|{\rm B}-{\rm L}        |>0$ would survive and not be erased at the weak scale by instanton effects. Thus baryogenesis at
$kT\sim 10^{10}-10^{15}~{\rm GeV}$ needs B-L violation and this is also needed to allow $m_\nu$ if neutrinos are Majorana particles.
The two effects could be related if baryogenesis arises from leptogenesis then converted into baryogenesis by instantons
\cite{buch}. The decays of heavy Majorana neutrinos (the heavy eigenstates of the see-saw mechanism) happen with violation of lepton number L, hence also of B-L and can well involve a sufficient amount of ¤CP violation. Recent results on neutrino masses are compatible with this elegant possibility. Thus the case
of baryogenesis through leptogenesis has been boosted by the recent results on neutrinos.

\section{Models of Neutrino Mixing}

After KamLAND, SNO and the upper limits on the absolute value of neutrino masses not too much hierarchy in the spectrum of neutrinos is indicated by experiments: 
\bea
r = \Delta m_{sol}^2/\Delta m_{atm}^2 \sim 1/30.\label{r}
\eea
Precisely $r=0.032^{+0.006}_{-0.005}$ at $3\sigma$'s  \cite{FogliIndication,MaltoniIndication}. Thus, for a hierarchical spectrum, $m_2/m_3 \sim \sqrt{r} \sim 0.2$, which is comparable to the Cabibbo angle $\lambda_C \sim 0.22$ or $\sqrt{m_{\mu}/m_{\tau}} \sim 0.24$. This suggests that the same hierarchy parameter (raised to powers with o(1) exponents) may apply for quark, charged lepton and neutrino mass matrices. This in turn indicates that, in the absence of some special dynamical reason, we do not expect quantities like $\theta_{13}$ or the deviation of  $\theta_{23}$ from its maximal value to be too small. Indeed it would be very important to know how small the mixing angle $\theta_{13}$  is and how close to maximal $\theta_{23}$ is. 

We see from Table(\ref{tab:data})  \cite{FogliIndication,MaltoniIndication} that within measurement errors
the observed neutrino mixing matrix is compatible with
the so called Tri-Bimaximal (TB) form \cite{hps}. The best measured neutrino mixing angle $\theta_{12}$ is just about 1$\sigma$ below the TB value $\tan^2{\theta_{12}}=1/2$, while the maximal value for $theta_{23}$ is well inside the 1-$\sigma$ interval and $theta_{13}$ is still compatible with zero(see table \ref{tab:data}).

The  TB mixing matrix (in a particular phase convention) is given by:
\begin{equation}
U_{TB}= \left(\matrix{
\dd\sqrt{\frac{2}{3}}&\dd\frac{1}{\sqrt 3}&0\cr
-\dd\frac{1}{\sqrt 6}&\dd\frac{1}{\sqrt 3}&-\dd\frac{1}{\sqrt 2}\cr
-\dd\frac{1}{\sqrt 6}&\dd\frac{1}{\sqrt 3}&\dd\frac{1}{\sqrt 2}}\right)~~~~~. 
\label{2}
\end{equation}

The TB mixing matrix suggests that mixing angles are independent of mass ratios (while for quark mixings relations like $\lambda_C^2\sim m_d/m_s$ are typical). In fact in the basis where charged lepton masses are 
diagonal, the effective neutrino mass matrix in the TB case is given by $m_{\nu}=U_{TB}\rm{diag}(m_1,m_2,m_3)U_{TB}^T$:
\begin{equation}
m_{\nu}=  \left[\frac{m_3}{2}M_3+\frac{m_2}{3}M_2+\frac{m_1}{6}M_1\right]~~~~~. 
\label{1k1}
\end{equation}
where:
\be
M_3=\left(\matrix{
0&0&0\cr
0&1&-1\cr
0&-1&1}\right),~~~~~
M_2=\left(\matrix{
1&1&1\cr
1&1&1\cr
1&1&1}\right),~~~~~
M_1=\left(\matrix{
4&-2&-2\cr
-2&1&1\cr
-2&1&1}\right).
\label{4k1}
\ee
The eigenvalues of $m_{\nu}$ are $m_1$, $m_2$, $m_3$ with eigenvectors $(-2,1,1)/\sqrt{6}$, $(1,1,1)/\sqrt{3}$ and $(0,1,-1)/\sqrt{2}$, respectively. The expression in eq.(\ref{1k1}) can be reproduced in models with sequential dominance or with form dominance, discussed by S. King and collaborators \cite{ski}. 

As we see the most general neutrino mass matrix corresponding to TB mixing, in the basis of diagonal charged leptons, is of the form:
\begin{equation}
m=\left(\matrix{
x&y&y\cr
y&x+v&y-v\cr
y&y-v&x+v}\right),
\label{gl21}
\end{equation}
This is a symmetric, 2-3 symmetric matrix with $a_{11}+a_{12}=a_{22}+a_{23}$.

Thus, one possibility is that one takes this coincidence seriously and considers models where TB mixing is a good first approximation.  In a series of papers \cite{TBA4,AFextra,AFmodular,AFL,afh,altverlin,altveram} it has been pointed out that a broken flavour symmetry based on the discrete
group $A_4$ appears to be particularly suitable to reproduce this specific mixing pattern in Leading Order (LO). Other
solutions based on alternative discrete or  continuous flavour groups have also been considered \cite{continuous,others,bmm}, but the $A_4$ models have a very economical and attractive structure, e.g. in terms of group representations and of field content. 

We recall that $A_4$, the group of even permutations of 4 objects, can be generated by the two elements
$S$ and $T$ obeying the relations (a "presentation" of the group):
\be
S^2=(ST)^3=T^3=1~~~.
\label{$A_4$}
\ee
The 12 elements of $A_4$  are obtained as:
$1$, $S$, $T$, $ST$, $TS$, $T^2$, $ST^2$, $STS$, $TST$, $T^2S$, $TST^2$, $T^2ST$.
The inequivalent irreducible representations of $A_4$ are 1, 1', 1" and 3. It is immediate to see that one-dimensional unitary representations are
given by:
\be
\begin{array}{lll}
1&S=1&T=1\\
1'&S=1&T=e^{\dd i 4 \pi/3}\equiv\omega^2\\
1''&S=1&T=e^{\dd i 2\pi/3}\equiv\omega \label{s$A_4$}
\end{array}
\ee
The three-dimensional unitary representation, in a basis
where the element $T$ is diagonal, is given by:
\be
T=\left(
\begin{array}{ccc}
1&0&0\\
0&\omega^2&0\\
0&0&\omega
\end{array}
\right),~~~~~~~~~~~~~~~~
S=\frac{1}{3}
\left(
\begin{array}{ccc}
-1&2&2\cr
2&-1&2\cr
2&2&-1
\end{array}
\right)~~~.
\label{ST}
\ee

Note that the generic mass matrix for TB mixing in eq.(\ref{gl21}) can be specified as the most general matrix that is invariant under $\mu-\tau$ symmetry, implemented by the unitary matrix  $A_{\mu \tau}$:
\be
A_{\mu \tau}=\left(
\begin{array}{ccc}
1&0&0\\
0&0&1\\
0&1&0
\end{array}
\right)
\label{Amutau}
\ee
and under the $S$ transformation:
\bea
m=SmS,~~~~~m=A_{\mu \tau}mA_{\mu \tau}~~\label{inv}
\eea
where S is given in eq.(\ref{ST}).
This observation plays a role in leading to $A_4$ as a candidate group for TB mixing, because $S$ is a matrix of $A_4$ (but  $A_{\mu \tau}$ is not and $\mu$-$\tau$ symmetry has to be separately implemented).

The flavour symmetry is broken by two triplets
$\varphi_S$ and $\varphi_T$ and by singlets $\xi$. 
All these fields are gauge singlets. The fields $\varphi_T$,
$\varphi_S$ and $\xi$ develop a VEV along the directions:
\bea
\langle \varphi_T \rangle=(v_T,0,0)~~~~~\langle \varphi_S\rangle=(v_S,v_S,v_S)~~~~~
\langle \xi \rangle=u. 
\label{align}
\eea 
A crucial part of all serious A4 models is the dynamical generation of this alignment in a natural way.
In most of the models $A_4$ is accompanied by additional flavour symmetries, either discrete like $Z_N$ or continuous like U(1), which are necessary to eliminate unwanted couplings, to ensure the needed vacuum alignment and to reproduce the observed mass hierarchies. In the leading approximation $A_4$ models lead to exact TB mixing.  Given the set of flavour symmetries and having specified the field content, the non leading corrections to the TB mixing arising from higher dimensional effective operators can be evaluated in a well defined expansion. In the absence of specific dynamical tricks, in a generic model, all the three mixing angles receive corrections of the same order of magnitude. Since the experimentally allowed departures of $\theta_{12}$ from the TB value $\sin^2{\theta_{12}}=1/3$ are small, at most of $\mathcal{O}(\lambda_C^2)$, with $\lambda_C$ the Cabibbo angle, it follows that both $\theta_{13}$ and the deviation of $\theta_{23}$ from the maximal value are expected in these models to also be at most of $\mathcal{O}(\lambda_C^2)$ (note that $\lambda_C$ is a convenient hierarchy parameter not only for quarks but also in the charged lepton sector with $m_\mu/m_\tau \sim0.06 \sim \lambda_C^2$ and $m_e/m_\mu \sim 0.005\sim\lambda_C^{3-4}$). A value of $\theta_{13} \sim \mathcal{O}(\lambda_C^2)$ is within the sensitivity of the experiments which are now in preparation and will take data in the near future. 

\section{$A_4$, quarks and GUT's}

Much attention  has been devoted to the question whether models with TB mixing in the neutrino sector can be  suitably extended to also successfully describe the observed pattern of quark mixings and masses and whether this more complete framework can be made compatible with (supersymmetric (SUSY)) SU(5) or SO(10) grand unification. Early attempts of extending models based on $A_4$ to quarks  \cite{ma1.5,AFmodular} and to construct grand unified versions \cite{maGUT} have not been  satisfactory, e.g. do not offer natural mechanisms for mass hierarchies and/or for the vacuum alignment. A direct extension of the $A_4$ model to quarks leads to the identity matrix for $V_{CKM}$ in the lowest approximation, which at first looks promising. But the corrections 
 to it turn out to be strongly constrained by the leptonic sector, because lepton mixings are nearly TB, and, in the simplest models, are proven to be too small to accommodate the observed quark mixing angles \cite{AFmodular}. Also, the quark classification adopted in these models is not compatible with $A_4$ commuting with SU(5) (in ref. \cite{KM} an $A_4$ model compatible with the Pati-Salam group SU(4)$\times$ SU(2)$_L \times$ SU(2)$_R$ has been presented). 
Due to this, larger discrete groups have been considered for the description of quarks  and for grand unified versions with approximate TB mixing in the lepton sector. A particularly appealing set of models is based on the discrete group $T'$, the double covering group of $A_4$ \cite{T'0}. In ref. \cite{T'} a viable description was obtained, i.e. in the leptonic sector the predictions of the $A_4$ model are reproduced, while the $T'$ symmetry plays an essential role for reproducing the pattern of quark mixing. But, again, the classification adopted in this model is not compatible with grand unification. Unified models based on the discrete groups $T'$ \cite{CM}, $S_4$ \cite{S4} and $\Delta(27)$  \cite{27} have been 
discussed. Several models using the smallest non-abelian symmetry 
$S_3$ (which is isomorphic to $D_3$) can also be found in the recent literature \cite{S3}.

As a result, the group $A_4$ was considered by most authors to be unsuitable to also describe quarks and to lead to a grand unified
description. We have recently shown \cite{afh} that this negative attitude
is not justified and that it is actually possible to construct a
viable model based on $A_4$ which leads to a grand
unified theory (GUT) of quarks and leptons with TB mixing
for leptons. At the same time our model offers an example of an
extra dimensional GUT in which a description of all fermion masses
and mixings is attempted. The model is natural, since most of the
small parameters in the observed pattern of masses and mixings as well
as the necessary vacuum alignment are  justified by the symmetries of
the model. The
formulation of SU(5) in extra dimensions has the usual advantages of
avoiding large Higgs representations to break SU(5) and of solving the
doublet-triplet splitting problem.  A see-saw realization
in terms of an $A_4$ triplet of right-handed neutrinos $\nu_R$ ensures the
correct ratio of light neutrino masses with respect to the GUT
scale. In our model extra dimensional effects directly
contribute to determine the flavour pattern, in that the two lightest
tenplets $T_1$ and $T_2$ are in the bulk (with a doubling $T_i$ and
$T'_i$, $i=1,2$ to ensure the correct zero mode spectrum), whereas the
pentaplets $F$ and $T_3$ are on the brane. The hierarchy of quark and
charged lepton masses and of quark mixings is determined by a
combination of extra dimensional suppression factors for the first two
generations and of the U(1) charges, while the neutrino mixing angles
derive from $A_4$. The choice of the transformation properties of the two
 Higgses $H_5$ and $H_{\bar{5}}$ is also crucial. They are chosen to transform 
as two different $A_4$ singlets
$1$ and $1'$. As a consequence, mass terms for the Higgs colour
triplets are  not directly allowed at all orders and their masses are
introduced by orbifolding, \`{a} la Kawamura \cite{5DSU5}. Finally, in this model, proton
decay is dominated by gauge vector boson exchange giving rise to
dimension six operators. Given the relatively large theoretical
 uncertainties, the decay rate is within the present
experimental limits. 
In conclusion, the model is shown to be directly compatible with approximate TB mixing for leptons 
as well as with a realistic pattern of fermion masses and of quark mixings in a SUSY SU(5) 
framework. 

\section{Bimaximal Mixing and S4}

Alternatively one can assume that the agreement of TB mixing with the data is accidental. Indeed there are many models that fit the data and yet TB mixing does not play a role in their architecture. For example, in ref.(\cite{alro}) there is a list of Grand Unified SO(10) models with fits to the  neutrino mixing angles that show good agreement with the data although most of them have no relation with TB mixing. Similarly for models based on $SU(5)\otimes U(1)$ \cite{review}. Another class of examples is found in ref.(\cite{seidl}. However, in most cases, for this type of models different mixing angles could also be accommodated by simply varying the fitted values of the parameters. Assuming that the agreement of TB mixing with the data is accidental, we observe that the present data do not exclude a larger value for $\theta_{13}$,$\theta_{13} \sim \mathcal{O}(\lambda_C)$, than generally implied by models with approximate TB mixing. In fact, two recent analyses of the available data lead to
$\sin^2{\theta_{13}}=0.016\pm0.010$ at 1$\sigma$ \cite{FogliIndication} and $\sin^2{\theta_{13}}=0.010^{+0.016}_{-0.011}$ at 1$\sigma$ \cite{MaltoniIndication}, which are compatible with both options. If experimentally it is found that $\theta_{13}$ is near its present upper bound, this could be interpreted as an indication that the agreement with the TB mixing is accidental. Then a scheme where instead the Bimaximal (BM) mixing is the correct first approximation could be relevant. 
The BM mixing matrix is given by:
\begin{equation}
U_{BM}= \left(\matrix{
\dd\frac{1}{\sqrt 2}&\dd-\frac{1}{\sqrt 2}&0\cr
\dd\frac{1}{2}&\dd\frac{1}{2}&-\dd\frac{1}{\sqrt 2}\cr
\dd\frac{1}{2}&\dd\frac{1}{2}&\dd\frac{1}{\sqrt 2}}\right)\;.
\label{21}
\end{equation}
In the BM scheme $\tan^2{\theta_{12}}= 1$, to be compared with the latest experimental
determination:  $\tan^2{\theta_{12}}= 0.45\pm 0.04$ (at $1\sigma$) \cite{FogliIndication,MaltoniIndication}, so that a rather large non leading correction is needed such that $\tan^2{\theta_{12}}$ is modified by terms of $\mathcal{O}(\lambda_C)$. This is in line with the well known empirical observation that $\theta_{12}+\lambda_C\sim \pi/4$, a relation known as quark-lepton complementarity \cite{compl}, or similarly $\theta_{12}+\sqrt{m_\mu/m_\tau} \sim \pi/4$. No compelling model leading, without parameter fixing, to the exact complementarity relation has been produced so far. Probably the exact complementarity relation is to be replaced with something like $\theta_{12}+\mathcal{O}(\lambda_C)\sim \pi/4$ or $\theta_{12}+\mathcal{O}(m_\mu/m_\tau)\sim \pi/4$ (which we could call "weak" complementarity), as in models where the large $\nu$ mixings arise from the diagonalisation of charged leptons.
Along this line of thought, we have used the expertise acquired with non Abelian finite flavour groups to construct a model \cite{S4us} based on the permutation group $S_4$ which naturally leads to the BM mixing at LO. We have adopted a supersymmetric formulation of the model in 4 space-time dimensions. The complete flavour group is $S_4\times Z_4 \times U(1)_{FN}$. In LO, the charged leptons are diagonal and hierarchical and the light neutrino mass matrix, after see-saw, leads to the exact BM mixing. The model is built in such a way that the dominant corrections to the BM mixing, from higher dimensional operators in the superpotential,  only arise from the charged lepton sector   and naturally inherit $\lambda_C$ as the relevant expansion parameter. As a result the mixing angles deviate from the BM values by terms of  $\mathcal{O}(\lambda_C)$ (at most), and weak complementarity holds. A crucial feature of the model is that only $\theta_{12}$ and $\theta_{13}$ are corrected by terms of $\mathcal{O}(\lambda_C)$ while $\theta_{23}$ is unchanged at this order (which is essential to make the model agree with the present data). 

\section{Conclusion}

In the last decade we have learnt a lot about neutrino masses and mixings.  A list of important conclusions have been reached. Neutrinos are not all massless but their masses are very small. Probably masses are small because neutrinos are Majorana particles
with masses inversely proportional to the large scale M of lepton number violation. It is quite remarkable that M is empirically not far from $M_{GUT}$, so that
neutrino masses fit well in the SUSY GUT picture. Also out of equilibrium decays with CP and L violation of heavy RH neutrinos can produce a B-L asymmetry, then converted near the weak scale by instantons into an amount of B asymmetry compatible with observations (baryogenesis via leptogenesis) \cite{buch}.  It has been established that neutrinos are not a significant component of dark matter in the Universe. We have also understood there there is no contradiction between large neutrino mixings and small quark mixings, even in the context of GUTÕs.  

This is a very impressive list of achievements. Coming to a detailed analysis of neutrino masses and mixings a very long collection of models have been formulated over the years. 
With continuous improvement of the data and more precise values of the mixing angles most of the models have been discarded by experiment. By now, besides the detailed knowledge of the entries of the $V_{CKM}$ matrix we also have a reasonable determination of the neutrino mixing matrix $U_{P-MNS}$. It is remarkable that neutrino and  quark mixings have such a different qualitative pattern. One could have imagined that neutrinos would bring a decisive boost towards the formulation of a comprehensive understanding of fermion masses and mixings. In reality it is frustrating that no real illumination was sparked on the problem of flavour. We can reproduce in many different ways the observations but we have not yet been able to single out a unique and convincing baseline for the understanding of fermion masses and mixings. In spite of many interesting ideas and the formulation of many elegant models the mysteries of the flavour structure of the three generations of fermions have not been much unveiled. 

\acknowledgments
I thank the Organizers of Les Rencontres, in particular my colleague at Roma Tre Mario Greco, for their kind invitation.

\end{document}